\documentclass[sigconf,nonacm]{acmart}

\setcopyright{none}
\usepackage{amsmath}
\usepackage{booktabs}
\usepackage{array}
\usepackage{multirow}
\usepackage{tabularx}
\usepackage[table]{xcolor}
\usepackage{graphicx}
\usepackage{microtype}
\usepackage{utfsym}
\usepackage{bbding}

\definecolor{oursblue}{HTML}{EAF2F8}

\title{TaskBridge: Technical Report for SIGIR 2026 eCommerce Challenge
}

\author{%
Zhiheng Fu$^{1}$,\quad
Yupeng Hu$^{1,\text{\Envelope}}$,\quad
Zixu Li$^{1,\dagger}$,\quad
Zhiwei Chen$^{1}$,\quad
Qinlei Huang$^{1}$,\quad
Shaokun Wang$^{2}$,\\
Han Liu$^{3}$,\quad
Fan Liu$^{4}$,\quad
Weili Guan$^{2}$,\quad
Liqiang Nie$^{2}$
\\[2mm]
$^{1}$Shandong University,\quad
$^{2}$Harbin Institute of Technology (Shenzhen),\\
$^{3}$City University of Hong Kong,\quad
$^{4}$Southeast University
}

\authorsaddresses{}

\AtBeginDocument{%
}

\thanks{%
  \textsuperscript{\ensuremath{\dagger}} Project Leader: Zixu Li.\quad
  \Envelope\,Corresponding Author: Yupeng Hu.
}

\begin{document}

\begin{abstract}
The Query Reformulation Task of the 2026 SIGIR Workshop on eCommerce assesses whether systems can reformulate task-oriented e-commerce queries that describe activities or goals into expressions better suited for product retrieval, and then retrieve the products required to complete the task through a standard BM25 pipeline. Unlike queries that directly specify products, task-oriented queries often encode product needs implicitly in activity-level semantics, creating a pronounced lexical gap between query terms and the product catalog. Moreover, the Automatic and Interactive protocols impose different requirements on the number of reformulations. A successful system must therefore preserve the original task intent while improving catalog-level retrievability, and select either a single query or multiple complementary queries according to the protocol.
We present TaskBridge, a catalog-anchored query reformulation framework for task-oriented e-commerce search. TaskBridge consists of the \textit{Catalog Semantic Anchoring} module, the \textit{Unified Five-Candidate Rewriting} module, and the \textit{Protocol-Aware Selection} module. Catalog Semantic Anchoring constructs a phrase lexicon and semantic anchors from the product corpus, providing task queries with searchable expressions grounded in catalog vocabulary. Unified Five-Candidate Rewriting generates multi-perspective candidates around the original query, preserving its task intent while covering different interpretations of the underlying product needs. Protocol-Aware Selection combines BM25 retrieval results, query--product reranking scores, and weighted Jaccard overlap to select one reformulation under the Automatic setting and up to four complementary reformulations under the Interactive setting. The selected queries are then executed independently using BM25 and converted into the six-column retrieval-result format required by the challenge. Experimental results show that TaskBridge effectively improves product retrieval for task-oriented queries. The multi-reformulation Interactive setting further exploits the complementarity among candidates, yielding additional gains in overall retrieval effectiveness.
TaskBridge achieved \textbf{first place} on the SIGIR 2026 eCommerce Challenge.

\end{abstract}

\maketitle
\thispagestyle{plain}


\section{Introduction}

Queries in e-commerce search do not always specify products explicitly. Users often express their needs through activities, scenarios, or desired outcomes. For example, ``make lasagna'' describes a cooking task without explicitly listing the baking dish, ingredients, or kitchenware required. Such task-oriented queries encode product needs in high-level intent, leaving no direct lexical correspondence between the query terms and product titles. Task-oriented query reformulation aims to preserve the original task intent while making relevant product expressions explicit, so that standard lexical retrieval can retrieve the products required to complete the task\cite{singh2012null,gollapudi2012structured,wu2017latent,shah2021task,sigirecom2026,zhang2026too,trecquerytask}. The central challenge is therefore not simple query expansion, but the transformation of a query into catalog-matchable product expressions without changing its task meaning.

This transformation involves two types of risk. When a query is under-reformulated, the original query may describe only the task goal and fail to reach the categories or entities that represent the required products in the catalog, making relevant products difficult to retrieve. When a query is over-reformulated, newly added terms may introduce products that the task does not require, weakening the original intent. Because product titles and category names in a catalog do not necessarily follow users' everyday expressions, generating seemingly plausible product terms with a language model alone does not guarantee a match to the actual catalog. Reformulation must therefore bridge the lexical gap between task intent and catalog expressions while controlling the catalog grounding and extent of the added content\cite{maji2019vocabulary,manchanda2019intent,hirsch2020query}.

More generally, automatic query-expansion research treats the mismatch between short keyword queries and underlying information needs as a central retrieval problem, and organizes solutions around expansion sources, term selection, and weighting strategies\cite{carpineto2012expansion,azad2019queryexpansion}. Neural information retrieval provides a complementary route by learning representations that bridge query and document vocabulary\cite{mitra2018neuralir}. Related work also explores multimodal state-space fusion~\cite{song3}. Sentence embeddings and dense dual-encoder retrieval represent text for semantic matching\cite{reimers2019sentencebert,karpukhin2020dpr}. Language-guided visual correspondence has also been studied~\cite{song4}, whereas neural passage reranking estimates the relevance of retrieved candidates\cite{nogueira2019reranking}. Recent advances in ranking calibration~\cite{RankVR} and adaptive composition strategies\cite{IMAGINE} further enhance retrieval quality through structure-aware optimization~\cite{song9}. More recently, Query2doc uses language-model-generated pseudo-documents to expand queries for sparse and dense retrieval\cite{wang2023query2doc}. And video-LLM research also studies fine-grained semantic understanding~\cite{song5}. These general approaches motivate the use of semantic anchors and a reranking proxy here, while TaskBridge keeps the final retrieval stage lexical and catalog-grounded.

This issue is particularly pronounced in lexical product retrieval. BM25 primarily ranks products according to term-level matches between the query and product titles\cite{robertsonwalker1994,robertsonzaragoza2009}; when a task-oriented query contains none of the product terms used in the catalog, BM25 cannot infer the required products from the task goal. Similar challenges arise in composed retrieval tasks, where multi-modal queries must be mapped to catalog items through semantic anchoring~\cite{OFFSET}. Conversely, adding a large number of product terms to the query may increase term overlap, but can also introduce products that are related to the task yet unnecessary. Efficient hybrid search techniques\cite{STABLE} and robust noise-mitigation strategies\cite{ERASE,REFINE} address related challenges in bridging semantic gaps while maintaining retrieval efficiency. Task-oriented query reformulation therefore requires a controllable mapping between task semantics and catalog vocabulary\cite{maji2019vocabulary,manchanda2019intent}.

The Query Reformulation Task at the 2026 SIGIR Workshop on eCommerce evaluates this problem under the Automatic and Interactive settings\cite{sigirecom2026,trecquerytask}. The Automatic setting requires one reformulation for each original query, whereas the Interactive setting permits up to four reformulations. These settings examine whether reformulated queries can support task-oriented product retrieval from single-query and multi-query perspectives, respectively, and provide an evaluation scenario for studying the transformation between task intent and product-catalog expressions.

To address this task, we propose TaskBridge, a catalog-anchored reformulation framework for task-oriented e-commerce queries. TaskBridge constructs a phrase lexicon and semantic anchors from product titles, generates candidates with the original query, retrieval feedback, and lexicon constraints, and selects queries according to the submission setting. It then submits the catalog-grounded reformulations to standard BM25 retrieval.

The main contributions of this paper are as follows:
\begin{itemize}
  \item We analyze the central challenge in task-oriented e-commerce query reformulation: preserving task intent while converting implicitly expressed product needs into vocabulary that matches catalog expressions;
  \item We propose TaskBridge, a catalog-anchored reformulation framework that integrates Catalog Semantic Anchoring, Unified Five-Candidate Rewriting, and Protocol-Aware Selection under the Automatic and Interactive settings;
  \item We validate the effectiveness of TaskBridge on the SIGIR eCom Query Reformulation benchmark. Under the Automatic setting, Task Completion NDCG@10, Essential Product Recall@100, and MAP reach $0.27$, $0.16$, and $0.11$, respectively; under the Interactive setting, they reach $0.31$, $0.18$, and $0.12$, respectively.
\end{itemize}

 \section{Method}
\begin{figure*}[ht]
\centering
\includegraphics[width=\linewidth]{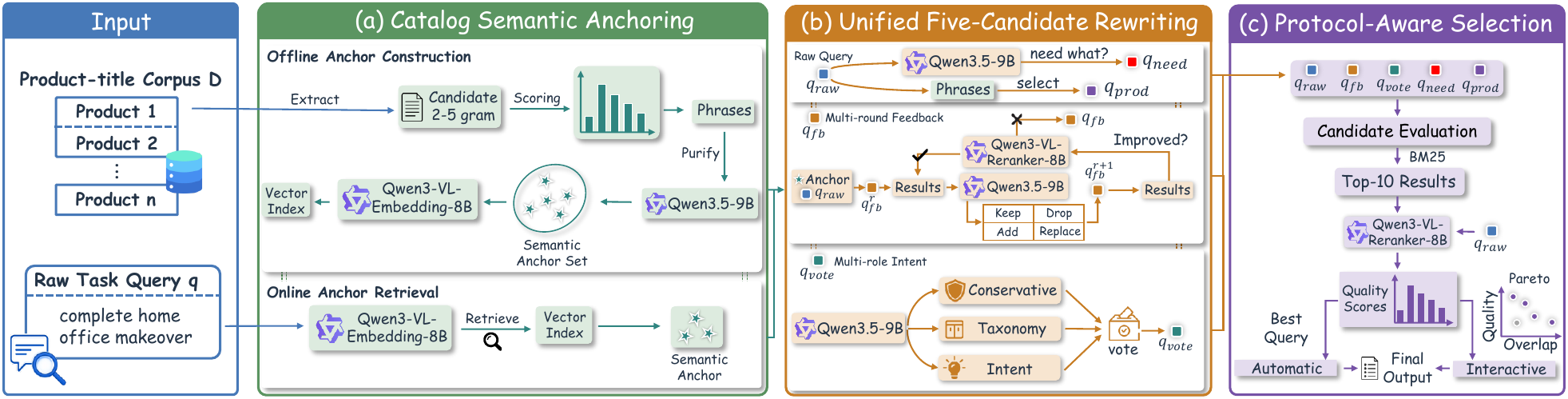}
\caption{Overall workflow of TaskBridge. The product corpus is first converted into catalog phrase resources. The two \textit{Phrases} nodes in the figure correspond to (a) and (b): the former is used to construct semantic anchors, whereas the latter supports constrained query reformulation. The original task query then produces five candidate reformulations. The Protocol-Aware Selection module retains the highest-quality candidate under a single-query budget and selects a candidate set that balances quality and result diversity under a multi-query budget. BM25 is finally executed independently for the selected queries.}
\label{fig:pipeline}
\end{figure*}

TaskBridge is an inference-time query reformulation framework for task-oriented e-commerce queries. It organizes catalog phrase resources, candidate query generation, and protocol-aware selection into a unified pipeline. First, it constructs catalog phrase resources for different downstream operations from the product corpus; it then generates candidate reformulations with complementary preferences around the original query; finally, it determines the submitted query set according to the Automatic or Interactive protocol. This design jointly considers task-intent preservation, catalog-vocabulary matching, and multi-query complementarity within a single selection framework. Figure~\ref{fig:pipeline} illustrates the overall architecture.

TaskBridge does not involve model training. By default, it uses Qwen3.5-9B~\cite{qwen35_9b_modelcard} to generate query reformulations and make role-prompted judgments, Qwen3-VL-Embedding-8B~\cite{li2026qwen3vl,bai2025qwen3vl,yang2025qwen3} to retrieve query-relevant semantic anchors, and Qwen3-VL-Reranker-8B~\cite{li2026qwen3vl,zhang2025qwen3embedding} to estimate the relevance between the original query and product titles. These models respectively support candidate construction, anchor retrieval, and selection-agent scoring, while the final product ranking follows the BM25 retrieval pipeline specified by the challenge.

\subsection{Problem Definition}

Given a task-oriented query $q$ and a product-title corpus $D$, TaskBridge aims to generate a set of queries suitable for lexical product retrieval and to return products relevant to the task intent through BM25.

BM25 is a probabilistic retrieval model based on term matching. Given a query $q$ and a product document $d$, its relevance score can be written as
\begin{equation}
\mathrm{BM25}(q,d)=\sum_{t\in q\cap d}\mathrm{IDF}(t)
\frac{f(t,d)(k_1+1)}{f(t,d)+k_1\left(1-b+b\frac{|d|}{\mathrm{avgdl}}\right)},
\end{equation}
where $f(t,d)$ is the frequency of term $t$ in document $d$, $|d|$ and $\mathrm{avgdl}$ are the document length and average document length, respectively, $k_1$ and $b$ control term-frequency saturation and length normalization, and $\mathrm{IDF}(t)$ is computed from the document frequency of the term. By increasing the contribution of rare matching terms while suppressing the term-frequency bias of long documents, BM25 ranks product titles~\cite{robertsonwalker1994,robertsonzaragoza2009}.

Based on the retrieval definition above, the task can be formalized as follows. Given an input pair $(q,D)$, the system generates one or more catalog-oriented reformulated queries and constructs a product ranking list for each reformulation. Let $F$ denote the reformulation function and $\mathcal{R}_{\mathrm{BM25}}$ the BM25 retrieval function. Then,
\begin{equation}
\widetilde{\mathcal{Q}}=F(q,D),\qquad
L_i=\mathcal{R}_{\mathrm{BM25}}(\widetilde{q}_i,D),
\quad \widetilde{q}_i\in\widetilde{\mathcal{Q}}.
\end{equation}

The reformulated queries should preserve the task intent expressed by $q$ as much as possible while narrowing the lexical gap between the task description and product titles. $L_i$ denotes the product ranking returned by the corresponding reformulated query.

\subsection{Catalog Semantic Anchoring}

Task-oriented queries typically describe activities, scenarios, or usage goals, whereas product catalogs primarily use category, entity, and attribute phrases~\cite{gollapudi2012structured,wu2017latent,maji2019vocabulary}. The Catalog Semantic Anchoring module constructs two complementary catalog phrase resources from the product corpus: a title phrase lexicon for constructing and purifying semantic anchors, and a corpus-derived phrase lexicon for providing candidate vocabulary for constrained query reformulation. Both resources share the product catalog as their semantic source, but differ in phrase-extraction scope and downstream use.

Let the product-title corpus be $D=\{d_1,\ldots,d_N\}$, and let $\mathcal{P}_{\mathrm{title}}(D)$ denote the set of candidate phrases extracted from the titles. The system filters two- to five-token phrases and ranks them using
\begin{equation}
s_{\mathrm{title}}(p)=(1+0.08n)\log(1+df(p))\log(1+cf(p)),
\end{equation}
where $n$ is the phrase length, $df(p)$ is the number of product documents containing phrase $p$, and $cf(p)$ is the number of occurrences of the phrase in the corpus. This produces the title phrase lexicon
\begin{equation}
\mathcal{V}_{\mathrm{title}}=\operatorname{TopK}_{p\in\mathcal{P}_{\mathrm{title}}(D)}s_{\mathrm{title}}(p).
\end{equation}

This lexicon provides stable category and entity expressions occurring in product titles and serves as the vocabulary source for semantic-anchor construction.

In parallel, the system extracts candidate phrases from the title, name, body, description, and category fields in the product corpus. Let $\mathcal{P}_{\mathrm{corp}}(D)$ denote the resulting candidate phrase set. These phrases form the corpus-derived phrase lexicon
\begin{equation}
\mathcal{V}_{\mathrm{corp}}=\operatorname{TopK}_{p\in\mathcal{P}_{\mathrm{corp}}(D)}s_{\mathrm{corp}}(p),
\end{equation}
where the candidate phrases consist of one- to three-token phrases and are scored using
\begin{equation}
s_{\mathrm{corp}}(p)=\bigl(1+0.25(n-1)\bigr)\log(1+df(p))\log(1+cf(p)).
\end{equation}

Here, $n$ is the phrase length, $df(p)$ is the number of product documents containing phrase $p$, and $cf(p)$ is the number of occurrences of the phrase in the product corpus. The corpus-derived lexicon supplies catalog-constrained phrase candidates for conservative need-aware and compact product-oriented reformulation, whereas semantic-anchor construction uses $\mathcal{V}_{\mathrm{title}}$. Subsequent equations use $\mathcal{V}_{\mathrm{title}}$ or $\mathcal{V}_{\mathrm{corp}}$ according to the operation.

Based on the title phrase lexicon, the system uses a generative model to convert candidate phrases into catalog semantic anchors. For each $p\in\mathcal{V}_{\mathrm{title}}$, the model determines whether it expresses a purchasable product category, object, or accessory and generates a concise normalized expression while preserving the product semantics. This process is written as
\begin{equation}
(\ell_p,\hat{p})=G_{\mathrm{purify}}(p),
\end{equation}
where $\ell_p$ is the semantic category of the candidate phrase and $\hat{p}$ is its normalized expression. To prevent scene descriptions, brand information, or title noise from being mistaken for product retrieval terms, the system retains only three types of expressions: product categories, product objects, and product accessories. It filters brands, scenes, marketing phrases, model fragments, and other non-product phrases. The semantic-anchor set is therefore defined as
\begin{equation}
\mathcal{A}=\{\hat{p}\mid p\in\mathcal{V}_{\mathrm{title}},\ \ell_p\in\mathcal{L}_{\mathrm{product}}\}.
\end{equation}

The purification process requires the model to return a structured category and normalized text for each input without generating additional explanations. This output is used directly to build the subsequent anchor index, excluding additional non-product terms and irrelevant details.

The system then constructs an embedding vector $\mathbf{e}(a_j)$ for each anchor $a_j\in\mathcal{A}$ and computes a query vector $\mathbf{e}(q)$ for the original query $q$. It retrieves query-relevant anchors using cosine similarity:
\begin{equation}
\operatorname{sim}(q,a_j)=
\frac{\mathbf{e}(q)^{\top}\mathbf{e}(a_j)}
{\lVert\mathbf{e}(q)\rVert_2\lVert\mathbf{e}(a_j)\rVert_2},
\qquad
\mathcal{A}(q)=\operatorname{TopK}_{a_j\in\mathcal{A}}\operatorname{sim}(q,a_j).
\end{equation}

Here, $\mathcal{A}(q)$ denotes the query-relevant semantic anchors. The title phrase lexicon provides stable catalog expressions for anchor construction, whereas the corpus-derived phrase lexicon provides broader vocabulary candidates for constrained reformulation. Both resources support candidate generation but neither replaces the final BM25 product retrieval. Related approaches disentangle complex semantics~\cite{TEMA,PAIR,MEDIAN} or construct intermediate-grained aggregations to bridge modality gaps, while our work focuses on task-to-catalog vocabulary mapping through phrase anchoring.

\subsection{Unified Five-Candidate Rewriting}

The Unified Five-Candidate Rewriting module takes the original query $q$, catalog phrase resources, and query-relevant semantic anchors as input and generates a candidate set with different preferences:
\begin{equation}
\mathcal{Q}(q)=\{q_{\mathrm{raw}},q_{\mathrm{fb}},q_{\mathrm{vote}},q_{\mathrm{need}},q_{\mathrm{prod}}\}.
\end{equation}

The candidate pool preserves the original intent, incorporates retrieval evidence, and applies different levels of catalog-vocabulary constraint. Its five candidates represent distinct reformulation hypotheses and are evaluated jointly by the Protocol-Aware Selection module.

\noindent\textbf{Original-query protection.}
The original-query candidate does not undergo generative reformulation and is defined as
\begin{equation}
q_{\mathrm{raw}}=q.
\end{equation}

It preserves the user's original wording and task context, providing an intent reference for the other candidates. When generative reformulation introduces unnecessary product terms or deviates from the original scenario, this candidate allows the selection process to return to the original expression.

\noindent\textbf{Multi-round feedback reformulation.}
Relevance feedback uses retrieval evidence to revise queries~\cite{harman1992relevance}; prior work also studies how to balance the original query with feedback information~\cite{lvzhai2009adaptive}. This candidate first generates an initial product expression from query-relevant semantic anchors and then iteratively revises it using product-title evidence returned by BM25. Let $q_{\mathrm{fb}}^{(0)}$ denote the initial reformulation, $E^{(r)}$ the product-title evidence returned by BM25 at round $r$, and $G_{\mathrm{fb}}$ the evidence-based feedback reformulator. Then,
\begin{equation}
E^{(r)}=\operatorname{TopK}_{d\in D}\operatorname{BM25}(q_{\mathrm{fb}}^{(r)},d),
\qquad
\widetilde{q}_{\mathrm{fb}}^{(r+1)}=G_{\mathrm{fb}}(q,q_{\mathrm{fb}}^{(r)},E^{(r)},\mathcal{A}(q)).
\end{equation}

The feedback reformulator proposes \texttt{Keep}, \texttt{Add}, \texttt{Drop}, and \texttt{Replace} edits for the current candidate. Applying these edits to the current candidate produces the candidate set
\begin{equation}
\mathcal{C}^{(r+1)}=
\operatorname{Edit}_{\{\mathrm{Keep,Add,Drop,Replace}\}}
\left(q_{\mathrm{fb}}^{(r)},\widetilde{q}_{\mathrm{fb}}^{(r+1)}\right).
\end{equation}

The system re-runs retrieval and quality estimation for every query in the candidate set and selects the candidate with the highest proxy quality:
\begin{equation}
\widehat{q}_{\mathrm{fb}}^{(r+1)}=\arg\max_{q'\in\mathcal{C}^{(r+1)}}\Phi(q').
\end{equation}

The update is accepted only when the aggregate reranking score of the new candidate exceeds that of the current candidate by more than a margin $\delta$:
\begin{equation}
q_{\mathrm{fb}}^{(r+1)}=
\begin{cases}
\widehat{q}_{\mathrm{fb}}^{(r+1)}, & \Phi(\widehat{q}_{\mathrm{fb}}^{(r+1)})>\Phi(q_{\mathrm{fb}}^{(r)})+\delta,\\
q_{\mathrm{fb}}^{(r)}, & \text{otherwise}.
\end{cases}
\end{equation}

Here, $\Phi(\cdot)$ denotes the candidate's reranking-quality proxy, and the candidate must not have been visited in any previous round. The update rule accepts a revision only when its proxy score exceeds the current score by more than $\delta$, thereby limiting changes during multi-round generation. Robustness to noisy signals during iterative refinement remains a challenge across retrieval paradigms\cite{INTENT,ConeSep,Air-know}.

\noindent\textbf{Multi-role intent reformulation.}
This candidate uses one generative model with three role prompts in a two-stage process~\cite{song7,song8}. The conservative role focuses on query drift, the taxonomy role determines whether a candidate expression belongs to a purchasable product category or object, and the intent role determines whether the scene description implies a clear shopping goal. In the first stage, the three roles independently judge whether the original query requires additional product expressions. Let $v_k(q)\in\{0,1\}$ denote the reformulation vote of role $k$. The system activates anchor selection only when the consensus reaches threshold $\tau$:
\begin{equation}
z(q)=\mathbb{I}\left[\sum_{k=1}^{K}v_k(q)\geq\tau\right].
\end{equation}

When $z(q)=1$, the three roles further select purchasable product categories or objects from $\mathcal{A}(q)$. Let $u_{k,j}(q)\in\{0,1\}$ indicate whether role $k$ supports candidate anchor $a_j$. The anchor support count and consensus-anchor set are respectively
\begin{equation}
u_j(q)=\sum_{k=1}^{K}u_{k,j}(q),
\qquad
\mathcal{A}_{\mathrm{vote}}(q)=
\{a_j\in\mathcal{A}(q)\mid u_j(q)\geq\tau_a\}.
\end{equation}

Here, $\tau_a$ is the anchor-level consensus threshold. The final candidate is written as
\begin{equation}
q_{\mathrm{vote}}=
\begin{cases}
\operatorname{Norm}\bigl(q\oplus\mathcal{A}_{\mathrm{vote}}(q)\bigr), & z(q)=1\ \text{and}\ \mathcal{A}_{\mathrm{vote}}(q)\neq\varnothing,\\
q, & \text{otherwise}.
\end{cases}
\end{equation}

Here, $\operatorname{Norm}$ performs deduplication and length control, and $\oplus$ denotes text concatenation. Role consensus separately determines whether reformulation is needed and which catalog terms to add. Hierarchical disambiguation strategies have been studied for ambiguous compositional queries\cite{ReTrack} provide related ways to refine query representations.

\noindent\textbf{Conservative need-aware reformulation.}
This candidate targets queries that already express a task scenario but lack product retrieval terms. The generator selects product phrases consistent with the task domain from the corpus-derived phrase lexicon and generates a bounded set of need terms. Let $N(q)$ denote the set of need terms and $P_{\mathrm{need}}(q)\subseteq\mathcal{V}_{\mathrm{corp}}$ the selected catalog phrases. The candidate can be written as
\begin{equation}
q_{\mathrm{need}}=\operatorname{Norm}\bigl(q\oplus N(q)\oplus P_{\mathrm{need}}(q)\bigr),
\end{equation}
where $\operatorname{Norm}$ denotes deduplication and length normalization, and $\oplus$ denotes text concatenation. This route constrains the number of newly added need terms, the total query length, and the number of selected phrases. These constraints define the route's conservative reformulation policy.

\noindent\textbf{Compact product-oriented reformulation.}
This candidate directly selects short product expressions from the corpus-derived phrase lexicon. Let $P_{\mathrm{prod}}(q)\subseteq\mathcal{V}_{\mathrm{corp}}$ denote the selected catalog phrases. Then,
\begin{equation}
q_{\mathrm{prod}}=\operatorname{Norm}\bigl(q\oplus P_{\mathrm{prod}}(q)\bigr).
\end{equation}

Compared with conservative need-aware reformulation, this route selects shorter category and entity phrases from the corpus-derived lexicon. The two lexicon-constrained candidates differ in reformulation extent and phrase-selection policy, and both are passed to the subsequent quality--diversity selection. Dual-path compositional networks\cite{HINT} and modification frequency-rarity balancing\cite{MELT} represent alternative strategies for controlling reformulation scope in retrieval tasks.

\subsection{Protocol-Aware Selection}

After candidate generation, the Protocol-Aware Selection module scores the five candidates using a unified retrieval proxy. For a candidate query $q_i$, BM25 first returns the top 100 products, denoted by
\begin{equation}
L_i=\{d_{i,1},\ldots,d_{i,100}\}.
\end{equation}

The reranker then judges query--product relevance for the top $K=10$ products using the original query $q$ and the corresponding product title. The resulting proxy scores the match between task intent and retrieved product evidence. Let $s_{i,r}$ denote the reranking score of the $r$-th product. We take the top $m=\min(5,K)$ scores and use logarithmic rank discounts, following the position-discounting principle of discounted cumulative gain~\cite{jarvelin2002cumulated}:
\begin{equation}
\omega(r)=\frac{1}{\log_2(r+1)},
\qquad
Q_i=\frac{\sum_{r=1}^{m}\omega(r)s_{i,r}}
{\sum_{r=1}^{m}\omega(r)}.
\end{equation}

Here, $Q_i$ is the proxy quality score of candidate $q_i$. This definition gives higher-ranked products greater influence on candidate scoring and connects BM25's lexical recall with the reranker's semantic judgment.

\noindent\textbf{Automatic single-candidate selection.}
The Automatic protocol permits one reformulation, so the selector returns the candidate with the highest quality proxy:
\begin{equation}
q_{\mathrm{auto}}=\arg\max_{q_i\in\mathcal{Q}(q)}Q_i.
\end{equation}

Automatic shares the candidate pool, BM25 retrieval, and reranking-quality estimation with Interactive selection.

\noindent\textbf{Interactive four-candidate selection.}
Interactive allows up to four reformulations to be submitted for each query. The selection objective must therefore consider both candidate quality and retrieval-result complementarity. This motivation relates to relevance--diversity trade-offs in MMR and proportional coverage~\cite{carbonell1998mmr,dang2012proportionality}, query-aspect coverage through reformulations~\cite{santos2010reformulations}, and diverse product-query reformulations~\cite{agrawal2023rlqr}. For candidates $i$ and $j$, let $r_i(d)$ denote the rank of product $d$ in $L_i$. If $d$ is absent from the list, its rank weight is set to zero. Define
\begin{equation}
\omega_i(d)=
\begin{cases}
\omega(r_i(d)), & d\in L_i,\\
0, & d\notin L_i,
\end{cases}
\end{equation}
then the weighted Jaccard overlap is
\begin{equation}
O(i,j)=\frac{\sum_{d\in L_i\cup L_j}\min(\omega_i(d),\omega_j(d))}
{\sum_{d\in L_i\cup L_j}\max(\omega_i(d),\omega_j(d))}.
\end{equation}

For any four-candidate subset $S\subseteq\mathcal{Q}(q)$ with $|S|=4$, the system computes its average quality and maximum pairwise overlap:
\begin{equation}
\overline{Q}(S)=\frac{1}{|S|}\sum_{q_i\in S}Q_i,
\qquad
O_{\max}(S)=\max_{q_i,q_j\in S,\ i\neq j}O(i,j).
\end{equation}

The selection objective is to maximize $\overline{Q}(S)$ and minimize $O_{\max}(S)$. If one subset has no lower average quality and no higher maximum overlap than another subset, with at least one objective strictly better, the latter is considered dominated and removed from the Pareto front. The system then selects one set from the non-dominated subsets using a deterministic rule that prioritizes quality and uses overlap as the secondary criterion, and outputs the four queries in descending order of their individual candidate quality.

In summary, Automatic and Interactive share the candidate pool, BM25 retrieval, and reranking-quality estimation, and can be understood uniformly as Pareto selection under different query budgets. Automatic prioritizes the candidate with the highest quality proxy under a single-query budget, whereas Interactive reduces retrieval-result overlap while maintaining set quality to preserve complementary reformulation directions.

\subsection{Final Retrieval Output}

After query selection, let the final set of retained reformulations be $\widehat{\mathcal{Q}}(q)$. The system independently executes BM25 for each reformulation and retains the corresponding product ranking list:
\begin{equation}
\widehat{L}(\widehat{q})=
\operatorname{Top100}_{d\in D}
\operatorname{BM25}(\widehat{q},d),
\qquad \widehat{q}\in\widehat{\mathcal{Q}}(q).
\end{equation}

Automatic produces one ranking list for each original query, whereas Interactive produces up to four independent ranking lists for each original query. Result lists corresponding to different reformulations are kept separate, allowing each reformulation to participate in evaluation as an independent retrieval direction. Finally, the system organizes the product rankings corresponding to the selected reformulations according to the applicable protocol and submits them in the evaluation format required by the challenge.

\section{Experiments}

\subsection{Challenge Setup}

TaskBridge is evaluated on the Query Reformulation Task of the 2026 SIGIR Workshop on eCommerce. The task focuses on task-oriented e-commerce queries and requires systems to transform users' descriptions of activities or goals into queries suitable for product retrieval, with reformulation quality assessed based on the retrieval results. The evaluation includes the Automatic and Interactive protocols: Automatic submits one reformulation for each original query, whereas Interactive permits up to four reformulations~\cite{sigirecom2026,trecquerytask}. TaskBridge uses the same product collection, task queries, catalog semantic resources, Unified Five-Candidate Rewriting module, and BM25 retriever under both protocols; the final difference lies in the number of reformulations that each protocol permits the system to retain.

\subsection{Implementation Details}

TaskBridge is an inference-time system with no training or fine-tuning. Candidate reformulations are generated by Qwen3.5-9B, semantic anchors are retrieved by Qwen3-VL-Embedding-8B, and relevance between queries and product titles is judged by Qwen3-VL-Reranker-8B~\cite{qwen35_9b_modelcard,li2026qwen3vl,li2026lora}. Candidate generation performs at most two rounds of retrieval feedback; semantic-anchor retrieval retains the top 50 candidates; multi-role voting uses three roles and requires at least two votes to support reformulation; and the two lexicon-constrained routes each select phrases from at most 80 candidates. Candidate selection uses BM25 results with a depth of 100, reranks the top 10 products, and aggregates the top 5 reranking scores. Automatic retains one candidate, whereas Interactive retains at most four. Each selected reformulation is then executed independently with BM25, retaining the top 100 products.

\subsection{Evaluation Metrics}

The challenge platform reports Task Completion NDCG@10, Essential Product Recall@100, and MAP. Task Completion NDCG@10 measures the extent to which the top ten products support task completion; Essential Product Recall@100 measures the coverage of essential products among the top 100 results; and MAP measures the average precision over the complete ranking list. These metrics reflect the quality of the result head, essential-product recall, and overall ranking performance, respectively~\cite{sigirecom2026,trecquerytask}.

\subsection{Performance Comparison}

Results are reported separately for the Open Phase and Closed Phase. Both tables use the same metric order: Task Completion NDCG@10, Recall@100, and MAP, and report results under both the Automatic and Interactive protocols.

\newif\ifprotocolgrouped
\protocolgroupedfalse

\ifprotocolgrouped
\begin{table}[t]
\centering
\small
\caption{Evaluation results of the official original-query BM25 baseline and \textit{TaskBridge (Ours)} in the Open Phase (Scheme B). The light-blue background marks our method; within each protocol, bold indicates the best result and underlining indicates the second-best result.}
\label{tab:open-phase-results}
\setlength{\tabcolsep}{3pt}
\begin{tabularx}{\columnwidth}{@{}p{0.34\columnwidth}*{3}{>{\centering\arraybackslash}X}@{}}
\toprule
Method & NDCG@10 & Recall@100 & MAP \\
\midrule
\multicolumn{4}{@{}l@{}}{\textbf{Automatic}} \\
\cmidrule(lr){1-4}
Original-query BM25 & \underline{0.42} & \textbf{0.30} & \underline{0.19} \\
\rowcolor{oursblue}
TaskBridge (Ours) & \textbf{0.49} & \underline{0.29} & \textbf{0.20} \\
\addlinespace
\multicolumn{4}{@{}l@{}}{\textbf{Interactive}} \\
\cmidrule(lr){1-4}
Original-query BM25 & \underline{0.42} & \underline{0.30} & \underline{0.19} \\
\rowcolor{oursblue}
TaskBridge (Ours) & \textbf{0.58} & \textbf{0.33} & \textbf{0.23} \\
\bottomrule
\end{tabularx}
\end{table}

\begin{table}[t]
\centering
\small
\caption{Evaluation results of \textit{TaskBridge (Ours)} in the Closed Phase (Scheme B). The light-blue background marks our method; as no results from other methods are available for this phase, no rankings are indicated.}
\label{tab:closed-phase-results}
\setlength{\tabcolsep}{3pt}
\begin{tabularx}{\columnwidth}{@{}p{0.32\columnwidth}p{0.18\columnwidth}*{3}{>{\centering\arraybackslash}X}@{}}
\toprule
Method & Protocol & NDCG@10 & Recall@100 & MAP \\
\midrule
\rowcolor{oursblue}
\multirow{2}{*}{TaskBridge (Ours)} & Automatic & 0.27 & 0.16 & 0.11 \\
\rowcolor{oursblue}
& Interactive & 0.31 & 0.18 & 0.12 \\
\bottomrule
\end{tabularx}
\end{table}

\else
\begin{table}[t]
\centering
\small
\caption{Evaluation results of the official original-query BM25 baseline and \textit{TaskBridge (Ours)} in the Open Phase. The light-blue background marks our method; within each protocol and metric, bold indicates the best result and underlining indicates the second-best result.}
\label{tab:open-phase-results}
\setlength{\tabcolsep}{2pt}
\begin{tabularx}{\columnwidth}{@{}>{\raggedright\arraybackslash}m{0.25\columnwidth}*{6}{>{\centering\arraybackslash}X}@{}}
\toprule
\multirow{3}{*}{Method} & \multicolumn{3}{c}{Automatic} & \multicolumn{3}{c}{Interactive} \\
\cmidrule(lr){2-4}\cmidrule(lr){5-7}
 & {\scriptsize\shortstack{NDCG\\@10}} & {\scriptsize\shortstack{Recall\\@100}} & MAP & {\scriptsize\shortstack{NDCG\\@10}} & {\scriptsize\shortstack{Recall\\@100}} & MAP \\
\midrule
Original-query & \underline{0.42} & \textbf{0.30} & \underline{0.19} & \underline{0.42} & \underline{0.30} & \underline{0.19} \\
\rowcolor{oursblue}
TaskBridge (Ours) & \textbf{0.49} & \underline{0.29} & \textbf{0.20} & \textbf{0.58} & \textbf{0.33} & \textbf{0.23} \\
\bottomrule
\end{tabularx}
\end{table}

As shown in Table~\ref{tab:open-phase-results}, the official Open Phase baseline applies BM25 directly to the original queries and therefore yields identical results under the two protocols. TaskBridge obtains scores of $0.49$, $0.29$, and $0.20$ under Automatic, and $0.58$, $0.33$, and $0.23$ under Interactive. Compared with the original-query baseline, TaskBridge's reformulations improve task-completion quality at the head of the results and overall ranking performance; Interactive further covers different product-expression directions by retaining multiple reformulations.

\begin{table}[t]
\centering
\small
\caption{Evaluation results of \textit{TaskBridge (Ours)} in the Closed Phase. The light-blue background marks our method.}
\label{tab:closed-phase-results}
\setlength{\tabcolsep}{2pt}
\begin{tabularx}{\columnwidth}{@{}>{\raggedright\arraybackslash}m{0.25\columnwidth}*{6}{>{\centering\arraybackslash}X}@{}}
\toprule
\multirow{3}{*}{Method} & \multicolumn{3}{c}{Automatic} & \multicolumn{3}{c}{Interactive} \\
\cmidrule(lr){2-4}\cmidrule(lr){5-7}
 & {\scriptsize\shortstack{NDCG\\@10}} & {\scriptsize\shortstack{Recall\\@100}} & MAP & {\scriptsize\shortstack{NDCG\\@10}} & {\scriptsize\shortstack{Recall\\@100}} & MAP \\
\midrule
\rowcolor{oursblue}
TaskBridge (Ours) & 0.27 & 0.16 & 0.11 & 0.31 & 0.18 & 0.12 \\
\bottomrule
\end{tabularx}
\end{table}
\fi

As shown in Table~\ref{tab:closed-phase-results}, TaskBridge obtains scores of $0.27$, $0.16$, and $0.11$ under Automatic, and $0.31$, $0.18$, and $0.12$ under Interactive in the Closed Phase. Within this phase, the Interactive metrics exceed the Automatic metrics by $0.04$, $0.02$, and $0.01$, respectively. Interactive retains up to four reformulations from the candidate pool, whereas Automatic retains one, so the two protocols use different submission budgets. The Open Phase and Closed Phase scores are platform records from separate phases and use different test sets; they are not directly comparable.

\section{Conclusion}
This paper presented TaskBridge, a catalog-anchored framework for task-oriented e-commerce query reformulation. By combining catalog semantic anchoring, five complementary rewriting strategies, and protocol-aware selection, TaskBridge maps implicit product needs to catalog expressions for BM25 retrieval. Interactive achieves higher scores than Automatic in both evaluation phases, suggesting that complementary reformulations can better support task-oriented product retrieval.

\bibliographystyle{unsrt}
\bibliography{references}

\end{document}